\documentclass[12pt]{article}
\usepackage{latexsym,epsfig,graphicx,amsmath,amssymb,amscd,multirow,chicago,psfrag,paralist,dsfont, undertilde, enumitem,dcolumn,bigstrut,lastpage}
\usepackage{subcaption, float}
\usepackage{wrapfig}
\usepackage{lscape,bm}
\usepackage{rotating}
\usepackage{lipsum}
\usepackage[american]{babel}

\newcommand{\thetavec}{{\boldsymbol{\theta}}}

\newcommand{\Xvec}{\boldsymbol{X}}

\newcommand{\Yvec}{\boldsymbol{Y}}

\newcommand{\xvec}{\boldsymbol{x}}
\newcommand{\svec}{\boldsymbol{s}}
\newcommand{\yvec}{\boldsymbol{y}}

\newcommand{\pvec}{\boldsymbol{p}}
\newcommand{\xvecj}{\boldsymbol{x}_j}
\newcommand{\xveci}{\boldsymbol{x}_i}

\begin{document}
	
\title{CASE STUDIES\\Design Strategies and Approximation Methods for High-Performance Computing Variability Management}
\author{Yueyao Wang{$^1$}, Li Xu{$^1$}, Yili Hong{$^1$}, Rong Pan{$^2$}, Tyler Chang{$^3$},
		\and
		 Thomas Lux{$^3$}, Jon Bernard{$^3$}, Layne Watson{$^3$}, and Kirk Cameron{$^3$}
		\\
		\\
		 \small {$^1$}Department of Statistics, Virginia Tech, Blacksburg, VA 24061
		 \\
		 \small {$^2$}School of Computing, Informatics, and Decision Systems Engineering,\\ \small Arizona State University, Tempe, AZ 85281		\\
		 \small {$^3$}Department of Computer Science, Virginia Tech, Blacksburg, VA 24060
	}
	
	\date{}
	\maketitle
\begin{abstract}
\textbf{Problem:} Performance variability management is an active research area in high-performance computing (HPC). In this paper, we focus on input/output (I/O) variability, which is a complicated function that is affected by many system factors. To study the performance variability, computer scientists often use grid-based designs (GBDs) which are equivalent to full factorial designs to collect I/O variability data, and use mathematical approximation methods to build a prediction model. Mathematical approximation models, as deterministic methods, could be biased particularly if extrapolations are needed. In statistics literature, space-filling designs (SFDs) and surrogate models such as Gaussian process (GP) are popular for data collection and building predictive models. The applicability of SFDs and surrogates in the HPC variability management setting, however, needs investigation. In this case study, we investigate their applicability in the HPC setting in terms of design efficiency, prediction accuracy, and scalability.

\textbf{Approach:} We first customize the existing SFDs so that they can be applied in the HPC setting. We conduct a comprehensive investigation of design strategies and the prediction ability of approximation methods. We use both synthetic data simulated from three test functions and the real data from the HPC setting. We then compare different methods in terms of design efficiency, prediction accuracy, and scalability.

\textbf{Results:} In our synthetic and real data analysis, GP with SFDs outperforms in most scenarios. With respect to the choice of approximation models, GP is recommended if the data are collected by SFDs. If data are collected using GBDs, both GP and Delaunay can be considered. With the best choice of approximation method, the performance of SFDs and GBD depends on the property of the underlying surface. For the cases in which SFDs perform better, the number of design points needed for SFDs is about half of or less than that of the GBD to achieve the same prediction accuracy. Although we observe that the GBD can also outperform SFDs for smooth underlying surface, GBD is not scalable to high-dimensional experimental regions. Therefore, SFDs that can be tailored to high dimension and non-smooth surface are recommended especially when large numbers of input factors need to be considered in the model. This paper has online supplementary materials.

\vspace{1ex}
		
\end{abstract}

\section{Problem Description} 

The computing scale and complexity in modern technologies and scientific areas make high-performance computing (HPC) increasingly important. Performance variability, however, is an important challenge in the research of HPC systems, which has been observed for a long time (e.g., \shortciteNP{giampapa2010experiences}, \citeNP{akkan2012stepping}, \shortciteNP{cameron2019moana}). High variability in HPC systems can lead to unstable system performance and potentially high energy costs. Therefore, variability management is crucial for system performance optimization. The performance variability is affected by many complicated interactions of factors in the system. In this study, we focus on input/output (I/O) performance variability. The relationship between system configurations (e.g., CPU frequency, file size, record size, the number of I/O threads, and I/O operation modes) and the I/O performance variability is of interest.

	One framework that has been used for HPC variability management involves three steps \shortcite{cameron2019moana}, which are data collection on performance variability for a set of system configurations, building an approximation model to make predictions for new configurations, and using the prediction to do optimization for future designs. Statistical research is usually involved in the data collection and prediction steps. Although there is vast research on designs for data collections and approximation models for predictions, the applicability of those statistical methods in the setting of HPC performance management needs investigation, which motivates us to do a case study based on the HPC performance data.
	
	In the data collection stage, researchers identify HPC system settings for which I/O throughput data should be collected. Computer scientists often use grid-based designs (GBDs) to collect data under numerous possible system configurations, when the number of factors is relatively small (\shortciteNP{cameron2019moana}, \shortciteNP{li2019mixture}). Note that the GBDs are equivalent to full factorial designs. In statistics literature, space-filling designs (SFDs) are often used, which assign design points far apart to fill the whole experiment region. In the prediction stage, an approximation model is built based on collected data for various system configurations. Mathematical approximation methods such as the linear Shepard's method \cite{shepard1968two} and Delaunay triangulation \cite{delaunay1934sphere} have been used. In statistics literature, Gaussian process (GP) models are popular for building approximation models.

From an HPC application point of view, there are several questions that need to be addressed. First, due to HPC system constraints, the design region could be irregular. For example, in our experiments, the file size needs to be larger than or equal to the record size, causing complexity in the experimental design. Therefore, existing SFDs need to be tailored so that they are suitable for the HPC setting. Second, it is desirable to demonstrate SFDs can collect data more efficiently than the GBD in a way that is accessible to computer scientists. Third, there is little research on the interaction between design strategies and approximation methods, especially in the setting of HPC variability management. However, it is possible that the prediction accuracy of a design may depend on the chosen approximation method.
	
Motivated by the needs in HPC performance variability management, we perform thorough comparisons between the prediction abilities of different approximation methods under different design strategies. We aim to recommend some efficient and scalable ways for computer scientists to collect performance data and provide a few practical guidelines in the HPC setting. Note that in this paper, the focus is on the modeling and analysis of HPC performance variability data collected from HPC experiments (i.e., not on the use of HPC to do statistical analysis).

	In literature, SFDs are widely used for experimental design when little information is known about the phenomenon to be studied. The uniform design proposed by \shortciteN{fang2000uniform} has the natural idea of placing design points uniformly in the experimental region. Latin hypercube designs \cite{mckay1979comparison} ensure good one-dimensional projection properties. Some design strategies are based on the distance measure, such as the maximin and minimax designs \cite{johnson1990minimax}. There are also several designs constructed based on the variants or combinations of the aforementioned designs, such as the maximin Latin hypercube design \cite{morris1995exploratory} and  the maximum projection design \cite{joseph2015maximum}.
	
	Non-regular and constrained design regions are quite common in industrial experiments and physical sciences. Some efforts have been made to allocate design points within such regions. \citeN{concad} introduce an algorithm to construct noncollapsing
	space-filling designs for bounded input regions. The design in \citeN{concad} can be applied to regions with constraints, but it can be time-consuming, especially when the design size is in hundreds or thousands. \citeN{lekivetz2015fast} propose the fast flexible space filling algorithm, which constructs designs based on hierarchical clustering. \shortciteN{pratola2017design} map the original dimension to a higher dimensional space to convert the geodesic distance to Euclidean distance. \citeN{golchi2015monte} adopt the idea of sampling from constrained distributions and propose a sequential constrained Monte Carlo algorithm to sample design points uniformly from the constrained input region. {\citeN{hung2010probability} propose probability-based Latin hypercube design for slid-rectangular region with the ability to achieve optimal design criteria.}
	
	When the physical experiment or the corresponding computer simulation model is complex and time-consuming, a surrogate model is needed to describe the underlying process. Many smooth techniques, such as response surface models, Kriging methods, kernel estimation, and neural networks, can be used to approximate the true surface. Response surface methodology, originally introduced by \citeN{box1951experimental}, is a traditional technique for modeling the response variables given input variables. Another commonly-used statistical approximation model is Gaussian process (GP) regression (e.g., \shortciteNP{sacks1989design}, \shortciteNP{currin1991bayesian}), which can generate a smooth surface and be capable of dealing with the heteroscedasticity \cite{goldberg1998regression} in the response variable. In the HPC community, mixture models have been used to study the multimodal behavior of the throughput distribution \shortcite{li2019mixture}. Some novel numerical techniques, including max box mesh, iterative box mesh, and Voronoi mesh methods for interpolation, are investigated by \shortciteN{lux2018novel}.
	
	In previous work, the SFDs and approximation models have been compared, separately, in terms of prediction performance. The prediction accuracy of SFDs is studied by \citeN{johnson2011empirical} with Gaussian process surrogates. Multiple approximation methods are compared under GBDs by \shortciteN{lux2018predictive} and \shortciteN{cameron2019moana}. Those compared methods include regression methods, such as multivariate adaptive regression spline model, support vector regression, multilayered perceptron regression, and some numerical methods, such as linear Shepard's method and Delaunay triangulation. However, little study has been done regarding the interaction between design strategies and approximation methods in the HPC performance setting. This paper aims to investigate the prediction ability of different kinds of design strategies and approximation methods as a case study. So, when a particular design strategy is given, HPC engineers can choose the best approximation method to achieve a higher prediction accuracy, or vice versa.

	The rest of the paper is presented as follows. Section~\ref{sec:datacollection} provides a detailed background of the motivating example and introduces the underlying problem that we are interested in. Sections~\ref{sec:design.strategies} and~\ref{sec:approximation.method} briefly summarize the SFDs and approximation methods that are under investigation in this paper. Section~\ref{sec:simulation.study} conducts synthetic data analyses where the performance of all combinations of the five designs and five approximation methods are compared under three test functions. In Section~\ref{sec:application}, an HPC variability experiment is introduced and real-world comparison results are presented. Section~\ref{sec:conclusion} provides conclusions of the comparisons, practical guidelines, and areas for future work.

\section{Data Collection and Preparation} \label{sec:datacollection}
	
We first introduce some notation for the HPC data. To define a general experiment process, let $\mathds{X}$ be a $d$-dimensional space of input factors with a set of constraints. Let $\Yvec$ be the random vector of the experiment outputs. The first step to start exploring the relationship between the input factors and the output is to efficiently allocate design points within $\mathds{X}$, at which the corresponding output vectors will be obtained. Suppose $\mathcal{D} = \left\{\xvec_1, \dots, \xvec_n \right\}$ is the set of selected design points, where $\xvec_i \in \mathds{X}$ is the $i$th design point. Let $y_i$ be the corresponding observed output at the $i$th design point. Then $\yvec = (y_1, \dots, y_n)$ is the collected output vector containing the observed outputs at all $\xvec_i \in \mathcal{D}$.

In the HPC data, there are four numeric factors, including one hardware factor: CPU frequency (GHz); and three application factors: the number of I/O threads, the file size (KB), and the record size (KB). For our experiments, we consider CPU frequencies in the range $[2.0, 3.5]$ GHz, numbers of I/O threads in the range $[1, 64]$, and file sizes and record sizes in the range $[4, 16384]$ KB.  Additionally, we have the constraint that for each system configuration, the file size must be greater than or equal to the record size, and the file size and record size must be of the form of $\sum_{l \in \mathcal{L}} 2^l$, where $\mathcal{L}$ is a set of positive integers. To make the design region uniform, we apply a log transformation to the file size and record size: $\textit{log file size} = \log_2 (\textit{file size})$, $\textit{log record size} = \log_2 (\textit{record size})$. Then the experiment region becomes $\mathds{X} = [2.0, 3.5] \times [1,64] \times [2,14] \times [2,14]$ with the constraint that $\textit{log file size} \geq \textit{log record size}$. In this application, we define the response of interest $\Yvec$ as the performance variability measurement (PVM). The PVM is given by the standard deviation of I/O throughputs (in the units of KB/s) at each input configuration (\shortciteNP{cameron2019moana}).
	
\begin{figure}
		\begin{center}
			\begin{tabular}{cc}
				\includegraphics[width=.47\textwidth]{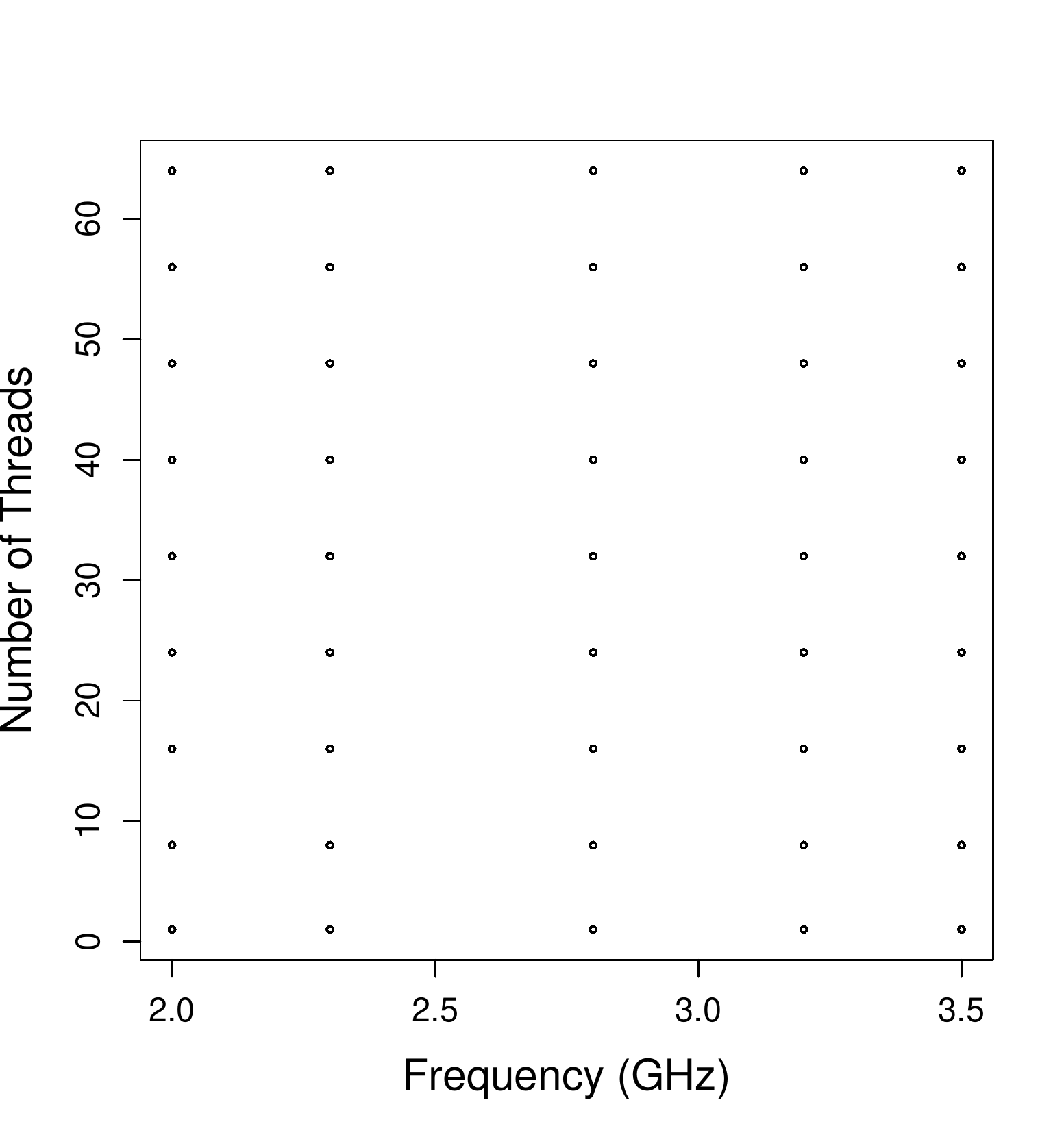}
				&
				\includegraphics[width=.47
				\textwidth]{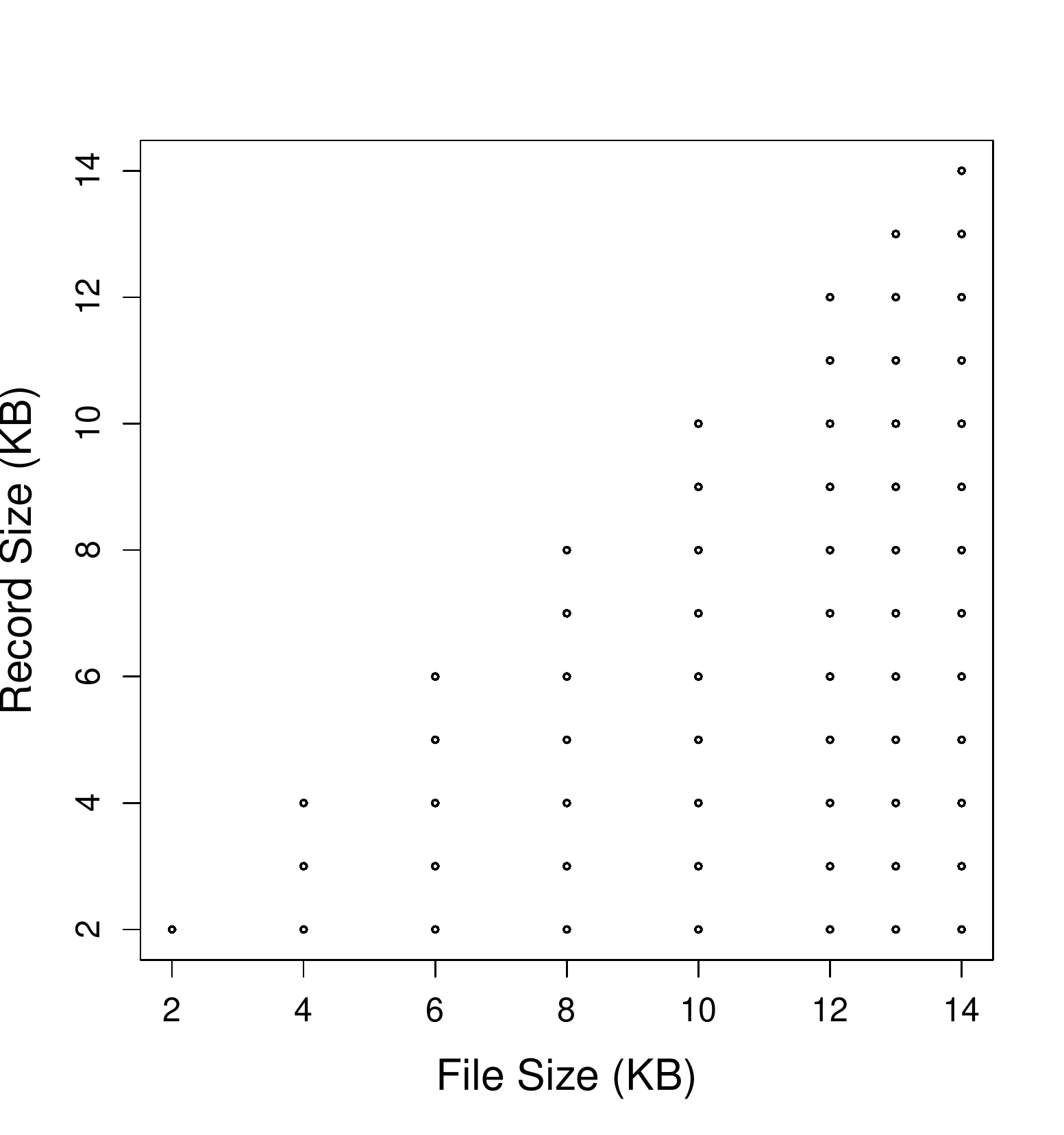}\\
				{\small (a) Frequency \& Number of Threads}& {\small (b) File Size \& Record Size}
			\end{tabular}
		\end{center}
		\caption{Two-dimensional projection grids of design points where the real data were collected.}
		\label{fig:real_grid_projection}
	\end{figure}

	To collect data for this application, the I/O throughput of hard disk is collected in a grid-based pattern, using the IOzone benchmark \citeyear{iozone} to produce the workload with each system setting. Each factor is divided into $k_i$ levels, $i = 1, \dots ,4$ and a configuration is obtained by taking a possible combination of levels in each factor. Figures~\ref{fig:real_grid_projection}(a) and~\ref{fig:real_grid_projection}(b) show the two-dimensional projections of the real 4-dimension grid space where data are collected. At each configuration, the IOzone benchmark is run multiple times and the throughputs are gathered as an HPC performance measurement. The configurations are collected under 6 IO operation modes: initial\_writers, rewriters, readers, re\_readers, random\_readers, random\_writers. In total, we have 2658 configurations and each configuration has 300 replications to capture the performance variability. Because the data collection procedure is time-consuming, we are interested in whether SFDs can select points more efficiently than the GBDs.

	After obtaining the data, another problem that we are interested in is the problem of accurately predicting HPC performance variability at a new configuration. With the collected dataset $\left\{ \mathcal{D}, \yvec \right\}$, we want to build models that describe the relationship between system factors and performance variability. Based on the previous literature and studies, we adopt both statistical models and numerical models in computer science to explore the underlying relationship.

	\begin{table}
		\caption{Illustration of the HPC performance variability data structure.}
		\centering
		\begin{tabular}{rrrrr}
			\hline\hline
			Frequency & No. of Threads & File Size & Record Size & PVM \\
			\hline
			2.0 &   1 & 2 & 2 & 17411.29 \\
			2.0 &   1 & 4 & 2 & 58014.19 \\
			2.0 &   1 & 4 & 3 & 46393.96 \\
			2.0 &   1 & 4 & 4 & 43238.60 \\
			2.0 &   1 & 6 & 2 & 49839.42 \\
			2.0 &   1 & 6 & 3 & 109721.24 \\
			\hline\hline
		\end{tabular}
		\label{tbl:data_str}
	\end{table}
	
\begin{figure}
		\begin{center}
			\begin{tabular}{cc}
				\includegraphics[width=.42\textwidth]{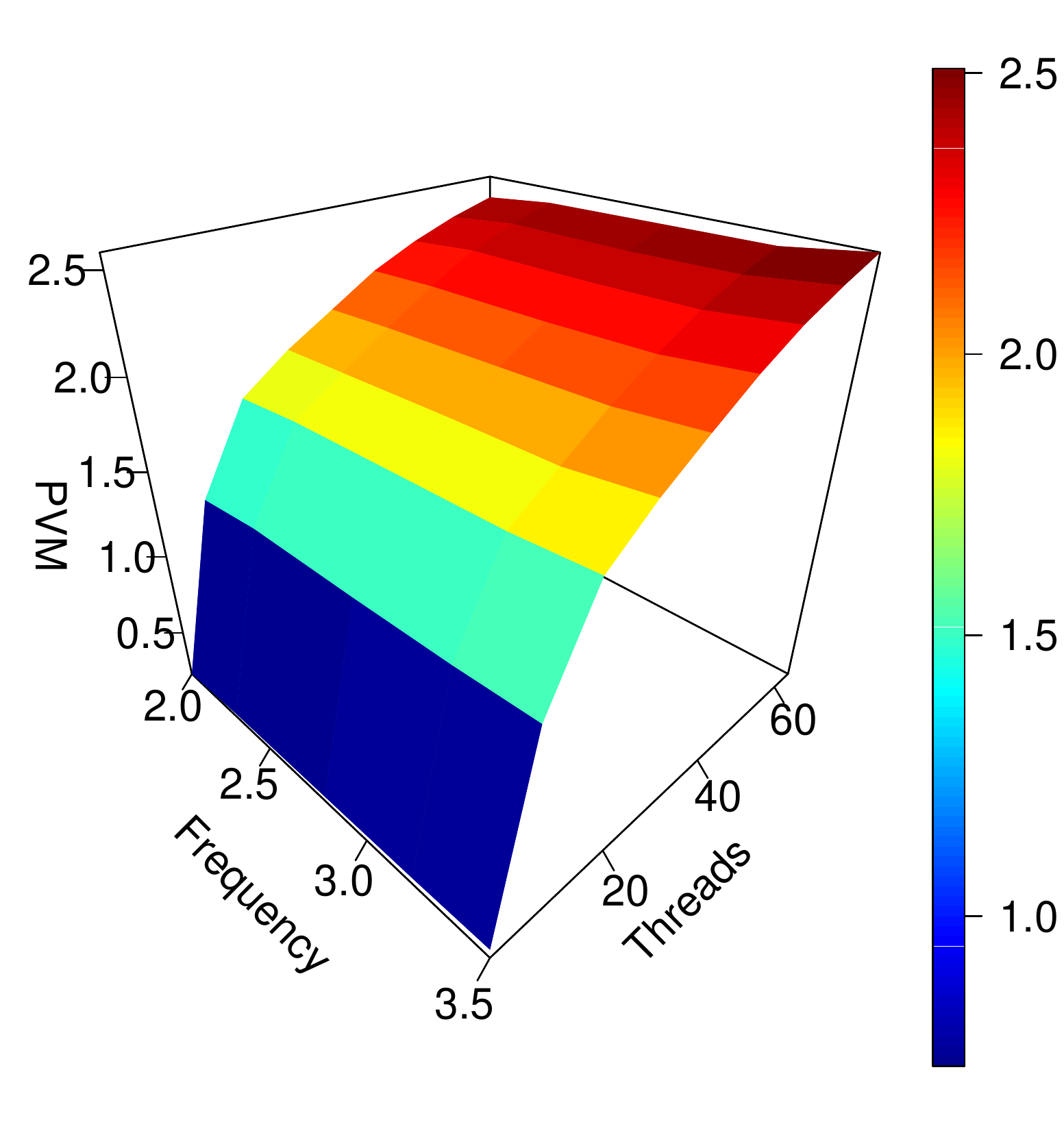}
				&
				\includegraphics[width=.42\textwidth]{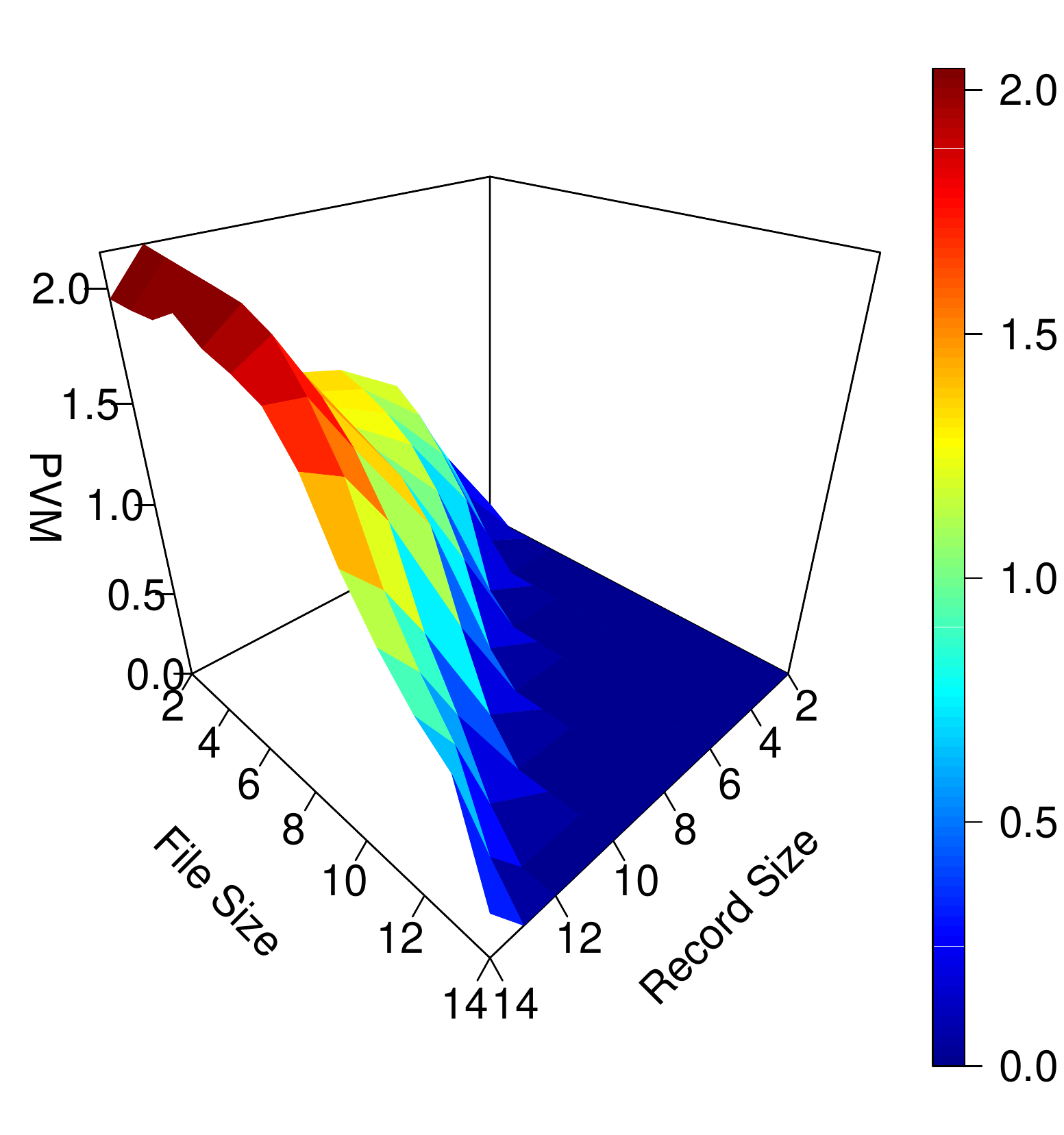}\\
				
				{\small (a) Frequency \& Threads} & {\small (b) File Size \& Record Size}
			\end{tabular}
		\end{center}
		\caption{3D surface plots of performance variability measure (PVM) on the magnitude of $10^6$ under two pairs of factors.}
		\label{fig:true_surface3d}
	\end{figure}

	An example of the data structure is presented in Table~\ref{tbl:data_str}. The 2D surface plots of the PVM under two pairs of factors are shown in Figure~\ref{fig:true_surface3d}. From the surface plot, we gain a rough idea of the relationship between response and input factors. In Figure~\ref{fig:true_surface3d}(a), we can see that the number of I/O threads and frequency have a positive relationship with the throughput variability; the larger the number of threads and frequency, the larger the PVM. Figure~\ref{fig:true_surface3d}(b) shows that when the file size and record size are closer to each other (i.e., near the boundary  of the constraint in the plot), the variability is relatively small. When the file size is much larger than the record size, the performance varies a lot. These relationships are consistent with our intuition.
	

\section{Analysis and Interpretation}
In this section, we conduct analyses on the effectiveness of designs and approximation methods in HPC setting. We first give brief descriptions on the design strategies and approximations in Sections~\ref{sec:design.strategies} and~\ref{sec:approximation.method}, respectively. We then examine the effectiveness of designs and approximation methods using synthetic data simulated from three different test functions. Finally, we study the designs and approximation methods using the real data from the HPC study.      	

\subsection{Design Strategies}\label{sec:design.strategies}	
	Space-filling designs (SFDs), as the name suggests, spread out design points evenly in the experiment region in order to gather information from the whole experiment region. SFDs can assign points based on distance measures or sampling strategies. One reason to propose SFDs as an alternative to GBDs is that spreading points evenly across the entire design region is ideal when prediction accuracy is our primary goal. This is because the prediction error at a particular point depends on its location relative to the design points. If a design allocates points only near the center of the experiment region, a large error may result in prediction for inputs on the boundary of the experiment region. For convenience, we first assume that the experiment region is a unit hypercube $\mathds{X} = \left[ 0,1\right]^d$. This assumption can easily be relaxed with linear transformations. This section introduces several common criteria for constructing SFDs, as well as a list of designs we will study in this paper.
	
\subsubsection{SFDs Constructed by Sampling Strategies}
	
	Several sampling strategies to construct SFDs are discussed in this section. An intuitive idea for a spread-out design is to scatter points uniformly in the design region. Designs built in this way are referred to as uniform designs (UDs), as described in \shortciteN{fang2000uniform}. UDs can gather a sufficient amount of information to explore the relationship between the response variable and the input factors with relatively small runs \cite{li2004}. UDs consider the whole design region equally important. However, when some portions of the domain are of more interest than others, stratified random sampling (SRS) can be used to enhance the design performance. Suppose $n$ design points are desired. The SRS partitions the design region $\mathds{X}$ into $s$ strata and in stratum $j$, $n_j$ points are selected based on a certain input distribution, where $j = 1, \dots, s$ and $\sum_{j=1}^s n_j = n$. The size and position of each stratum depends on different experiment scenarios. When we know that some input factors are important to the response, we also want the design points' projections on to those factors to be spread out. This can be achieved by the Latin hypercube design (LHD) \cite{mckay1979comparison}, which is a popular SFD and has been combined with various other designs. To construct an LHD of size $n$, the range of each input factor is equally divided into $n$ intervals $\left[0, 1/n\right), \dots, \left[(n-1)/{n}, 1 \right]$. For each of the $d$ coordinates, exactly one design point projection is sampled from each interval. In this way, it can be guaranteed that the design points are spread out across the range of each input factor. One favorable property of LHD is that any lower dimension projection of an LHD is also an LHD. Although LHDs have good properties for projection, it is not guaranteed that the design points will be spread out evenly in the entire experiment region. For example, it is possible that all design points could be located along the diagonal of the $d$-dimensional unit hypercube. To overcome this drawback, Latin hypercube sampling is often used together with other designs such as maximin design \cite{johnson1990minimax} or maximum projection design \cite{joseph2015maximum}. These SFDs are based on a distance metric, which will be discussed in the next section.
\subsubsection{SFDs Based on Distance Measures}

	In this section, design strategies based on distance measures and metrics are briefly introduced. One idea for SFDs is that no point in the experiment region $\mathds{X}$ should be too far from its nearest neighbor in $\mathcal{D}$. Let $$d(\xveci, \xvecj) = \left( \sum_{k=1}^{d} |x_{ik} - x_{jk}|^s\right)^{1/s}$$ be the distance metric. Usually, the Euclidean distance, (i.e., $s=2$) is used. For an arbitrary point $\xvec$ in $\mathds{X}$, we determine its closest design point $\xvec_i$ and the minimum distance $\min_i d(\xvec, \xvec_i)$. To guarantee that no point is too far from the design points, we choose the point $\xvec \in \mathds{X}$ with the maximum distance to its closest design point, which is $\max_{\xvec \in \mathds{X}} \min_i d(\xvec, \xvec_i)$. Then we find the design to minimize this distance,
	$$\min_{\mathcal{D}} \max_{\xvec \in \mathds{X}} \min_i d(\xvec, \xvec_i),$$
	which is called the minimax distance design \cite{johnson1990minimax}. The second way to spread out points in $\mathcal{D}$ is to allocate the design points as far apart as possible. This can be realized by finding the design that maximizes the minimum distance between two design points, which is the criterion of the maximin distance design,
	\begin{equation*}
	\max_{\mathcal{D}} \min_{i,j}d(\xvec_i, \xvec_j).
	\end{equation*}
	The maximin distance design ensures that the design points are spread as far apart from each other as possible in the full dimension but does not guarantee that the design is space filling for each projection on a subspace. \citeN{morris1995exploratory} propose the maximin Latin hypercube (MmLh) design, which incorporates the structure of an LHD with the maximin design. The criterion is
	\begin{equation}
	\min_{\mathcal{D}} \left[ \sum_{i=1}^{n-1} \sum_{j = i+1}^{n}\frac{1}{d^m(\xvec_i, \xvec_j)}  \right]^{1/m}.
	\label{eq:mmlhd_criterion}
	\end{equation}
	
	The MmLh design ensures good projection properties when projecting into one dimension, but fails to consider the projection onto other subdimensions. \citeN{joseph2015maximum} propose a maximum projection design (MaxPro) that ensures good space filling on all subspaces. It considers a weighted Euclidean distance
	\begin{equation}
	d(\xvec_i, \xvec_j, \thetavec) = \left[ \sum_{k=1}^{d} \theta_k (x_{ik} - x_{jk})^2\right]^{1/2},
	\label{eq:weighted_d}
	\end{equation}
	where $\thetavec = (\theta_1, \dots, \theta_d)$ is a vector of weight factors. Let $\theta_k = 1$ for those factors that construct the subspace and $\theta_k = 0$ for other factors, then \eqref{eq:weighted_d} calculates the distance between $\xvec_i$ and $\xvec_j$ after projection into subspaces. The criterion of maximum projection design can be modified based on \eqref{eq:mmlhd_criterion} as	
$$\min_{\mathcal{D}} \sum_{i=1}^{n-1} \sum_{j = i+1}^{n}\frac{1}{d^m(\xvec_i, \xvec_j, \thetavec)}.$$
	Here, the weight factors satisfy $\sum_{k=1}^d \theta_k = 1$. To properly choose the weight factor $\thetavec$, \citeN{joseph2015maximum} adopt the Bayesian framework. A prior distribution is assigned to $\thetavec$ and the expected value of the objective function is minimized. The criterion can be further simplified with uniform prior and $m = 2d$:
	\begin{equation}
	\min_{\mathcal{D}} \sum_{i=1}^{n-1} \sum_{j = i+1}^n \frac{1}{\prod_{k=1 }^{d} \left( x_{ik} - x_{jk}\right)^2}.
	\label{eq:maxpro_criterion}
	\end{equation}
	
	It is clear from \eqref{eq:maxpro_criterion} that any two design points in $\mathcal{D}$ can not have the same value in any dimension. Otherwise, $x_{ik} - x_{jk} = 0$, which will cause the objective function to be infinite. The criterion automatically guarantees that the design also has the LHD property.
	
	Another design strategy, the maximum entropy design is also included in this section, although generally it is considered to be a model-based design. The maximum entropy design \cite{shewry1987maximum} maximizes the negative entropy function, which captures the amount of information gained from an experiment. For a random vector $\Xvec$ with support $\mathds{X}$ and density function $f(\Xvec)$, the entropy is defined as $H(\Xvec) = -\text{E}_{\Xvec}[\log f\left(\Xvec\right)]=-\int_{\mathds{X}} f(\xvec)\log[f(\xvec)]d\xvec$. The information gained is $I(\Xvec) = -H(\Xvec)$. We want the experimental design that causes the largest change in the information, which is equivalent to maximizing the entropy:
	$$\max_{\mathcal{D}}H(\Xvec).$$
	This can be simplified as $\max_{\mathcal{D}}\log|\Sigma_n|$ if we assume that the underlying surface is a Gaussian process, where $\Sigma_n$ is the variance-covariance matrix of $\mathcal{D}$. Then the construction of the maximum entropy design depends on the choice of the correlation function. In this study, we adopt the format that is implemented in the R package \textit{DiceDesign} \cite{dicedesign}, where the correlation function is $r(\xvec_i,\xvec_j) = 1 - 1.5d/a - 0.5 (d/a)^3$ if $d>a$ and $0$ otherwise. Here, $d$ is the Euclidean distance between the two points ($\xvec_i$ and $\xvec_j$) and $a$ denotes the range of the variogram.
	
\subsubsection{Customizing SFDs for HPC Setting}\label{sec:taylorsfds}
	Four of the above designs are considered in this paper. They are uniform sampling, maximum Latin hypercube design (MnLh), maximum entropy design (MaxEnt), and maximum projection LHD (MaxPro). In the HPC performance variability modeling application, the additional constraint is that the file size needs to be larger than or equal to the record size. To deal with the constraint in the real application, we want to use an approach that can tailor all different designs. So above designs must be adjusted in order to accommodate the constrained region. In this study, a rejection sampling strategy, as a simple way that can be easily applied across different design strategies, is used. Suppose we would like to generate a design with size $n$. If points in the initial size $n$ design fail to meet the constraint, then we generate designs with size $2n$, $3n$, $\dots$, until there are at least $n$ points in the design that satisfy the constraint. Then we randomly select $n$ points from the large design, which satisfy our constraint.

Note that this idea is different from rejection sampling, because we reject points that are obtained from a specific optimization problem. After discarding the points that do not meet the constraint, the remaining design points may not hold all the original properties of that type of design. However, this is a uniform, simple approach that can be applied to any type of design. We do not need to solve extra optimization problems or alter the existing sampling algorithm.
	
Besides tailoring SFDs for the constraint in the HPC problem, we also need to adjust designs to include boundary points so that we will not have bad extrapolation problems when we using the data to build numerical prediction models. The details are explained in Section~\ref{sec:delaunay}. In this study, we use central composite designs (CCDs) to augment origin SFDs. A CCD consists of a factorial (fractional factorial) design with factors of two levels. That is, the vertices of the experiment region, a set of center points, and a set of axial points. We add the vertex points, the center point of the axis, and the center point of the hypercube to the SFDs. If the experiment region is irregular, then the augmented points that do not meet the constraint are excluded.

\subsection{Approximation Methods}\label{sec:approximation.method}
	
	To conduct the prediction accuracy comparison, we need to select several representative surrogates to approximate the underlying function. The approximation methods often used in both computer science and statistics are briefly introduced here. In applied mathematics, numerical methods are often used to approximate the true surface. In this section, two numerical methods are used, which are Delaunay triangulation and Linear Shepard's method.
\subsubsection{Delaunay Triangulation Interpolation}\label{sec:delaunay}
	
	Delaunay triangulation interpolation is a numerical method that approximates the underlying function $f: \mathds{R}^d \rightarrow \mathds{R}$ based on the values $f(\pvec)$ for a given vertex set $\mathcal{P}$ and the corresponding Delaunay triangulation. A Delaunay triangulation \cite{delaunay1934sphere} is given by any triangulation that satisfies the Delaunay properties. Suppose $\mathcal{P} = \left\{ \pvec_1, \dots, \pvec_n \right\}$ is a set of $n$ points in $\mathds{R}^d$. Then a $d$-dimensional triangulation of $\mathcal{P}$, denoted $\textrm{T}({\mathcal{P}})$, is a set of $d$-simplices that satisfies the following criteria: 1) the vertex set of $\textrm{T}(\mathcal{P})$ is $\mathcal{P}$; 2) the union of all simplices in $\textrm{T}(\mathcal{P})$ is the convex hull of $\mathcal{P}$, denoted as $\textrm{CH}(\mathcal{P})$; and 3) the $d$-simplices are disjoint besides their common boundaries (vertices, edges and facets). A Delaunay triangulation, denoted as $\textrm{DT}(\mathcal{P})$, is the geometric dual of the Voronoi diagram \cite{watson1981computing}. A Voronoi diagram divides $\mathds{R}^d$ into $n$ regions with each containing one Delaunay vertex in $\mathcal{P}$, such that all points within each region are closer to the vertex in their region than to any other vertex. The Delaunay triangulation is given by connecting all vertices whose Voronoi cells share a boundary with an edge.
	
	Let $\xvec \in \textrm{CH}(\mathcal{P})$ be an interpolation point and $S \in \textrm{DT}(\mathcal{P})$ be the simplex with vertices $\svec_1, \dots, \svec_{d+1}$ that contains $\xvec$. One finds the weights $w_1,\dots, w_{d+1}$ that satisfy $\sum_{i=1}^{d+1} w_i= 1$, $w_i \geq 0$, for $i = 1, \dots, d+1$ and $\xvec = \sum_{i=1}^{d+1} w_i \svec_i$. Then the estimated function value $\widehat{f}_{\textrm{DT}}(\xvec)$ based on $\textrm{DT}(\mathcal{P})$ is
	$$\widehat{f}_{\textrm{DT}}(\xvec) = w_1 f(\svec_1) +  w_2 f(\svec_2) + \dots + w_{d+1}f(s_{d+1}).$$
	
	In this paper, the Fortran 2003 package DELAUNAYSPARSE \shortcite{chang2020algorithm} is used to perform the interpolations. In order to achieve computational efficiency, the algorithm in DELAUNAYSPARSE only computes a necessary, sparse subset of the Delaunay triangulation given pre-specified interpolation points \shortcite{chang2018polynomial}.
	
	In the HPC variability management application, in order to compute the predictions, we need to approximate the response values for a test set based on the Delaunay triangulation $\textrm{DT}(\mathcal{D})$ of the proposed design $\mathcal{D}$. However, the points in the test set might not always fall inside $\textrm{CH}(\mathcal{D})$, which results in an extrapolation problem instead of an interpolation problem. DELAUNAYSPARSE can handle extrapolation problems by projecting each test point onto $\textrm{CH}(\mathcal{D})$ when the test point is close to $\textrm{CH}(\mathcal{D})$, but this solution can perform badly if the test point is far outside of $\textrm{CH}(\mathcal{D})$. Since we do not know whether a test point is inside $\textrm{CH}(\mathcal{D})$ beforehand, necessary adjustments of the design strategies need to be made to avoid bad extrapolation problems.
	
	In order to avoid bad extrapolation problems when using the Delaunay method to approximate the true surface, we need to augment the proposed SFDs so that the convex hull of each augmented SFD covers the entire experimental region. In this way, wherever a test point falls in the experimental region, it always results in an interpolation problem for the Delaunay method. To realize this idea, intuitively one can augment each SFD with those boundary points. In this study, we choose to use CCD to augment the proposed SFDs.
\subsubsection{Linear Shepard Method}
	Shepard's method is a form of inverse distance weighting, originally proposed by \citeN{shepard1968two}. It is an interpolation method based on the weighted average of basis functions, each centered on a point in the data set $\mathcal{P} = \{\pvec_1, \dots, \pvec_n\}$, where the weight is calculated in terms of inverse distance from the interpolation point to points in $\mathcal{P}$. Several variations of Shepard's method are available such as, quadratic and cubic Shepard's method. However, in our application, linear Shepard's method is used for the sake of efficiency. Given the same setting as in Section~\ref{sec:delaunay}, the original Shepard approximation of $f(\xvec)$ at point $\xvec$ is:
	\begin{equation*}
	\widehat{f}(\xvec) = \frac{\sum_{i = 1}^n W_i(\xvec)f(\pvec_i) }{\sum_{i = 1}^n W_i(\xvec)},
	\label{eq:org_lsp}
	\end{equation*}
	where $W_i(\xvec) = 1/{\Vert \xvec - \pvec_i\Vert_2^2}$. This weight is nonzero for all the data points even for those points that are far away from the interpolation point $\xvec$. In order to achieve better approximation performance, a modified linear Shepard's method considers only the local points within an $R$-sphere of $\xvec$ and replaces the original $f(\pvec_i)$ with a linear approximation function $B(\pvec_i)$. The modified linear Shepard's method has the form
	\begin{equation*}
	\widehat{f}(\xvec) = \frac{\sum_{i = 1}^n W_i(\xvec)B(\pvec_i) }{\sum_{i = 1}^n W_i(\xvec)},
	\label{eq:modified_lsp}
	\end{equation*}
	where the modified version weights are given by $$W_i(\xvec) = \left[\frac{\max\big(0, R_i-d(\xvec, \pvec_i)\big)}{R_i d(\xvec, \pvec_i)}\right]^2.$$
	Here, $R_i$ is the radius of a sphere centered at $\xvec_i$ that reflects influence scope of $\xvec_i$. The Fortran package SHEPPACK \shortcite{thacker2010algorithm} is used in our study to perform the linear Shepard interpolation.
	
	Besides the numerical methods, three statistical models are considered in this study: response surface methodology (RSM), multivariate adaptive regression splines (MARS) and Gaussian processes (GPs). {Compared to numerical models, one advantage of statistical models is that one can quantify the prediction uncertainty in a relatively easy manner.}
	
\subsubsection{Response Surface Methodology}	
	
	Response surface methodology (RSM) is a method that investigates the relationship between a response variable and input factors through design experiments. It is a traditional method for studying computer experiments (e.g., \citeNP{myers1999response}, \shortciteNP{myers2004response}). Low-order polynomial models such as first- or second-degree polynomial models are commonly used in applications. In this paper, we use a second-order model to approximate the true response surface.
	
\subsubsection{Multivariate Adaptive Regression Splines}
	Multivariate adaptive regression splines (MARS) were introduced by \citeN{friedman1991multivariate}. They are nonparametric regression models formed by spline basis product expansion. MARS can automatically capture nonlinearity and interaction effects. The MARS model is given by:
	\begin{equation*}
	\widehat{f}(x) = \sum_{m=1}^M c_m B_m(x),
	\end{equation*}
	where $B_m(x)$ are the basis functions. These basis functions can be constants, hinge functions, or products of hinge functions, where each hinge function is of the form $\max(0, x- c)$ or $\max(0, c - x)$, where $c$ is a constant. As a flexible nonparametric model, MARS tends to overfit without pruning. The algorithm for constructing MARS model is based on a modification of recursive partition trees that requires a forward and backward pass. In the forward pass, the MARS model is initialized as a constant valued function (whose value is the intercept), then basis functions are gradually added to the model until the maximum number of terms is reached or the loss in sum of squared residuals is small. Next, a backward pass is used to prune the model based on the generalized cross validation (GCV) criterion, which is a trade off between goodness-of-fit and model complexity. In this study, we use the \textit{earth} package in R to build the MARS model \cite{earthpackage}. A grid of hyper-parameter: the number of maximum terms in the model and the maximum degree of interactions, is used to train the model in order to the select model with the highest R-squared value.
	
\subsubsection{Gaussian Process}
	Gaussian process (GP) interpolation is a commonly used approximation method in computer experiments. GP is often defined as a stochastic process where every finite collection of $n$ observations follows a multivariate normal (MVN) distribution. A GP is determined by its mean function $\mu(\xvec)$ and its covariance function $C(\xvec, \xvec')$. When a zero-mean GP is assumed, it can be completely determined by the covariance function, which is also referred to as the kernel function. There are several common kernel functions, including Gaussian $C(\xvec, \xvec') = \exp[-{(\xvec - \xvec')^2}/{\theta}]$ and Mat\'ern kernels.
	Let $\Yvec_n = \left\{y_1, \dots, y_n \right\}$ be the $n$ observations at the proposed design points $\Xvec_n = \left\{\xvec_1, \dots, \xvec_n \right\}$ and $\Yvec \sim \textrm{N}_n(0, \Sigma_n)$, where $\Sigma_n$ is the covariance matrix with covariance elements $C(\xvec, \xvec')$. Then for an arbitrary point $\xvec$, the value of $Y(\xvec)$ given the design and corresponding observations can be obtained by the conditional distribution $Y(\xvec)|\left\{\Yvec_n, \Xvec_n \right\}$. This can be calculated based on the conditional distribution of MVN:
	\begin{equation*}
	Y(\xvec)|\left\{\Yvec_n, \Xvec_n \right\} \sim \textrm{N}[\mu(\xvec), \sigma^2(\xvec)],
	\end{equation*}
	where $\mu(\xvec) = \Sigma(\xvec, \Xvec_n)\Sigma_n^{-1}\Yvec_n$ and $\sigma^2(\xvec) = \Sigma(\xvec, \xvec) - \Sigma(\xvec, \Xvec_n)\Sigma_n^{-1}\Sigma(\Xvec_n,\xvec)$. Here, $\Sigma(\xvec, \Xvec_n)$ is a $1 \times n$ matrix with elements $C(\xvec, \xvec_1), \dots, C(\xvec, \xvec_n)$. It is obvious that the mean function is a linear combination of $\Yvec_n$ while the covariance function does not involve information of observations. A maximum likelihood estimator can be used for parameter estimation. We use the R package \textit{laGP} \cite{laGPpackage} to implement the local Gaussian process approximation. In the local GP model, if one wants to predict at $\xvec$, instead of using the whole design $\mathcal{D}$, a subset of $\mathcal{D}$ close to $\xvec$ is selected sequentially to increase the computing speed. {Note that when building the GP model, both the input variables and the response are normalized to range $[0,1]$.}

\subsection{Synthetic Data Analyses}\label{sec:simulation.study}
	
	In order to investigate the performance of each design strategy and approximation method combination, we conduct synthetic data analyses using three test functions. Specifically, if we denote $y_i$ as the true response for the $i$th design point and $\hat{y}_i$ is the predicted value, $i = 1, \dots, n$, we compare the root mean squared error (RMSE) $\sqrt{\sum_{i = 1}^n(\hat{y}_i - y_i)^2/n}$ and the mean absolute percentage error (MAPE) $\sum_{i = 1}^n \left(|\hat{y}_i - y_i|/y_i\right)/n$ of the predictions for each combination with each test function.
	
	We have two goals when conducting the synthetic data analyses. First, we want to perform simulations with test functions that are representative of the HPC performance. Since this application has four input factors, we choose two test functions that have four input variables each, and for each of these test functions, we apply the same linear constraint function as in the HPC performance variance problem. Second, we would like to explore the prediction behavior and computing time of GBD and SFD with a high-dimension test function. So the eight-dimensional Borehole function is adopted to illustrate the design performance when the experiment region is of high dimension.

\subsubsection{Test Functions}

The Colville function is a four-dimensional function with the formula:
\begin{align*}
f(\xvec) &= 100(x_1^2 - x_2)^2 + (x_1 - 1)^2 + (x_3 - 1)^2 + 90 (x_3^2 - x_4)^2 + \\
&10.1 ((x_2 - 1)^2 + (x_4 - 1)^2 ) + 19.8 (x_2 - 1)(x_4 - 1),
\end{align*}
where $\xvec = (x_1, \dots, x_4)$.
The Colville function's input domain is $x_i \in [-10, 10]$, $i = 1,\dots, 4$. We apply the constraint $x_3 \geq x_4$ to the input domain to emulate the HPC performance variability problem.

	The Friedman function was proposed by \citeN{friedman1983multidimensional}. It is a five-dimensional function, and in our study, we map the first four variables to our experiment region and fix the last variable $x_5$. The function is:
	\begin{equation*}
	f(\xvec) = 10 \sin(\pi x_1 x_2) + 20 (x_3 - 0.5)^2 +10 x_4 + 5x_5,
	\end{equation*}	
	where $\xvec = (x_1, \dots, x_4)$. This function's domain is the unit hypercube: $x_i \in [0, 1]$, $i = 1,\dots,4$ with $x_5 = 0.5$. The same constraint as with Colville function was applied here.

The Borehole function is an eight-dimensional function that models water flow rate through a borehole. Let $\xvec = \left(r_w, r, T_u, H_u, T_l, H_l, L, K_w \right)$ be the input variables, then the water flow rate is:
	\begin{equation*}
	f(\xvec) = \frac{2\pi T_u(H_u - H_l)}{\log(r/r_w)\left(1+ \frac{T_u}{T_l} + \frac{2LT_u}{\log(r/r_w)K_wr_w^2}\right)}.
	\end{equation*}
	The input ranges are listed in Table~\ref{tbl:boreholerange}. For this test function, no constraint is applied because the goal of this test function is to investigate the high-dimensional performance of the design strategy and the approximation method combinations.
	
	\begin{table}
		\centering
		\caption{Ranges of parameters for the Borehole function.}
		\begin{tabular}{crr}
			\hline\hline
			Variable & Minimum & Maximum \\
			\hline
			$r_w$ & 0.05 & 0.15 \\
			$r$ & 100 & 50000 \\
			$T_u$ & 63070 & 115600 \\
			$H_u$ & 990 & 1110 \\
			$T_l$ & 63.1 & 116 \\
			$H_l$ & 700 & 820 \\
			$L$ & 1120 & 1680 \\
			$K_w$ & 9855 & 12045 \\
			\hline\hline
		\end{tabular}
		\label{tbl:boreholerange}
	\end{table}
\subsubsection{Comparison Procedures} \label{sec:simulation.procedure}
	Using the above test functions, we want to compare the prediction accuracy of GBDs with that of proposed SFDs for each test function using various design sizes and approximation methods. Since the GBD is built by selecting $n$ levels on each factor and then enumerate all possible combinations of $d$ factors, the design size needs to be $n^d$. In a real application, it is possible that the numbers of levels at each factor are different. However, since our test functions are continuous functions, we assume that if we can take $n$ levels on one factor, we can also take the same number of levels on all other factors, which means we only consider the fine ``regular grid" in the synthetic data analyses. After deciding the size of GBD, we can generate SFDs correspondingly. The simulation procedure is as follow:
	
	\begin{enumerate}[itemsep=0mm]
		\item Choose a test function. Uniformly select $n_g$  random points within the input region as the test set $g$.	
		\item For $n = 3,\ldots,7,$
		\begin{enumerate}
			\item Create a GBD $\mathcal{D}_g$ by choosing $n$ points in each dimension and expanding into a grid via the Cartesian product. Exclude design points that do not satisfy the constraint and denote the size of GBD as $N_g$.
			\item \label{enulst:sfd} To generate the SFDs, create designs $\mathcal{D}_{maximin}$, $\mathcal{D}_{maxpro}$, $\mathcal{D}_{maxent}$, and $\mathcal{D}_{uniform}$ each of size $N = N_g - n_a$, where $n_a$ is the size of the augmented design.
			\item \label{enulst:evl_test} For each test function, find the corresponding values of points in the above designs.
			\item \label{enulst:pred_error} Use five approximation methods to generate five predictive surfaces for each design:
			\begin{enumerate}
				\item Linear regression: use backward selection to determine the second-order linear regression model with minimum Bayesian information criterion (BIC).
				\item Delaunay triangulation: use the DELAUNAYSPARSE package to build the model.
				\item Linear Shepard: use the SHEPPACK package to build the model.
				\item MARS: use cross validation to tune the MARS model on a hyper-parameter grid, and select the model with the highest R-squared value.
				\item Gaussian process: use the separable Gaussian kernel with a nugget effect included in the model.
			\end{enumerate}
			\item Repeat Steps \ref{enulst:evl_test} - \ref{enulst:pred_error} $B$ times.
			\item Compute RMSE and MAPE over repetitions for each SFD and approximation method.
		\end{enumerate}	
	\end{enumerate}
	
	Although in Section~\ref{sec:design.strategies} we introduced that the SFDs are obtained by solving different optimization problems, those optimization problems often do not have analytic solutions if large number of points are desired in a high-dimensional experiment region. Numerical algorithms are usually used to obtain SFDs, which lead to non-unique solutions for a certain type of SFD. In order to understand the overall prediction performance for a certain type of SFD, we repeat the step of generating designs and making predictions for $B$ times. In the above procedures, the test set size $n_g$ and the repetition number $B$ is changed according to the dimension of the test function and the approximation method. For relatively smooth test functions, (e.g., Friedman function) or stable approximation methods, (e.g., Delaunay and MARS), we do not require a large number of replications or a large test set to obtain a stable and representative result. However for other models, such as using the linear Shepard's method under a non-smooth test function, we need to increase the replication number in order to get a reliable result. The summary of the test sizes $n_g$ and replication numbers $B$ are listed in Table~\ref{tbl:sim_pars_summary}.
	
	For the Borehole function, since the dimension is relatively high, it is difficult to compute the prediction performance for a series of GBDs within a reasonable time and with reasonable computational resources. Therefore, we skip the step of generating multiple GBDs of the same size as each SFD. Instead, we consider one GBD of size $3^8 = 6561$ GBD and compare its performance with other SFDs varying sizes from $\left\{500, 600, \dots, 2000\right\}$.
	
\begin{figure}
\centering
\includegraphics[width=1\textwidth]{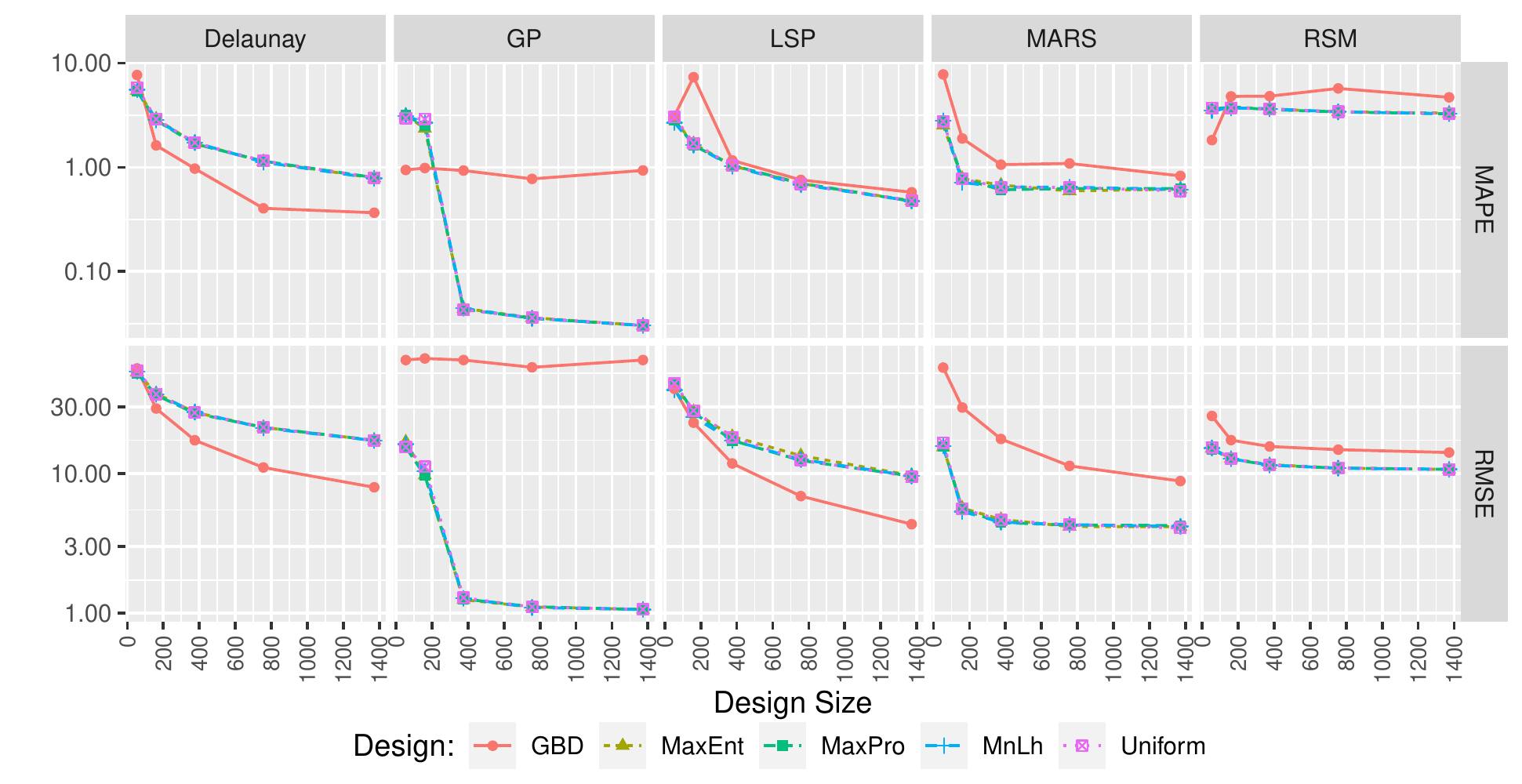}
\caption{Plot of RMSE and MAPE as functions of design size under the Colville test function. RMSE is on the magnitude of $10^4$.} \label{fig:colsimulation}
\end{figure}

\begin{figure}
\centering
			\includegraphics[width=1.0\textwidth]{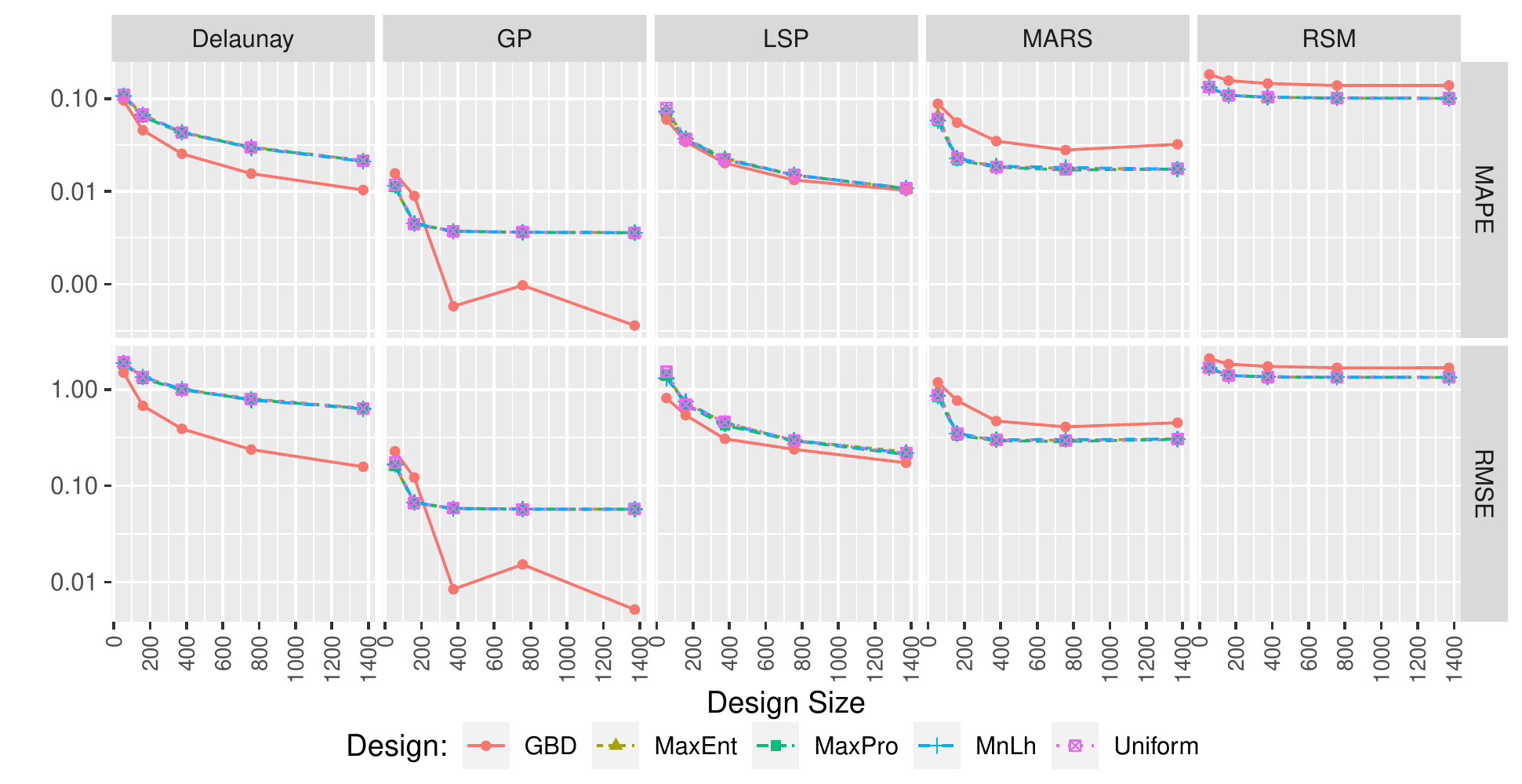}
			
			\caption{Plot of RMSE and MAPE as functions of design size under the Friedman test function.}
			\label{fig:friedsimulation}
\end{figure}

\begin{figure}
\centering
\includegraphics[width=1.0\textwidth]{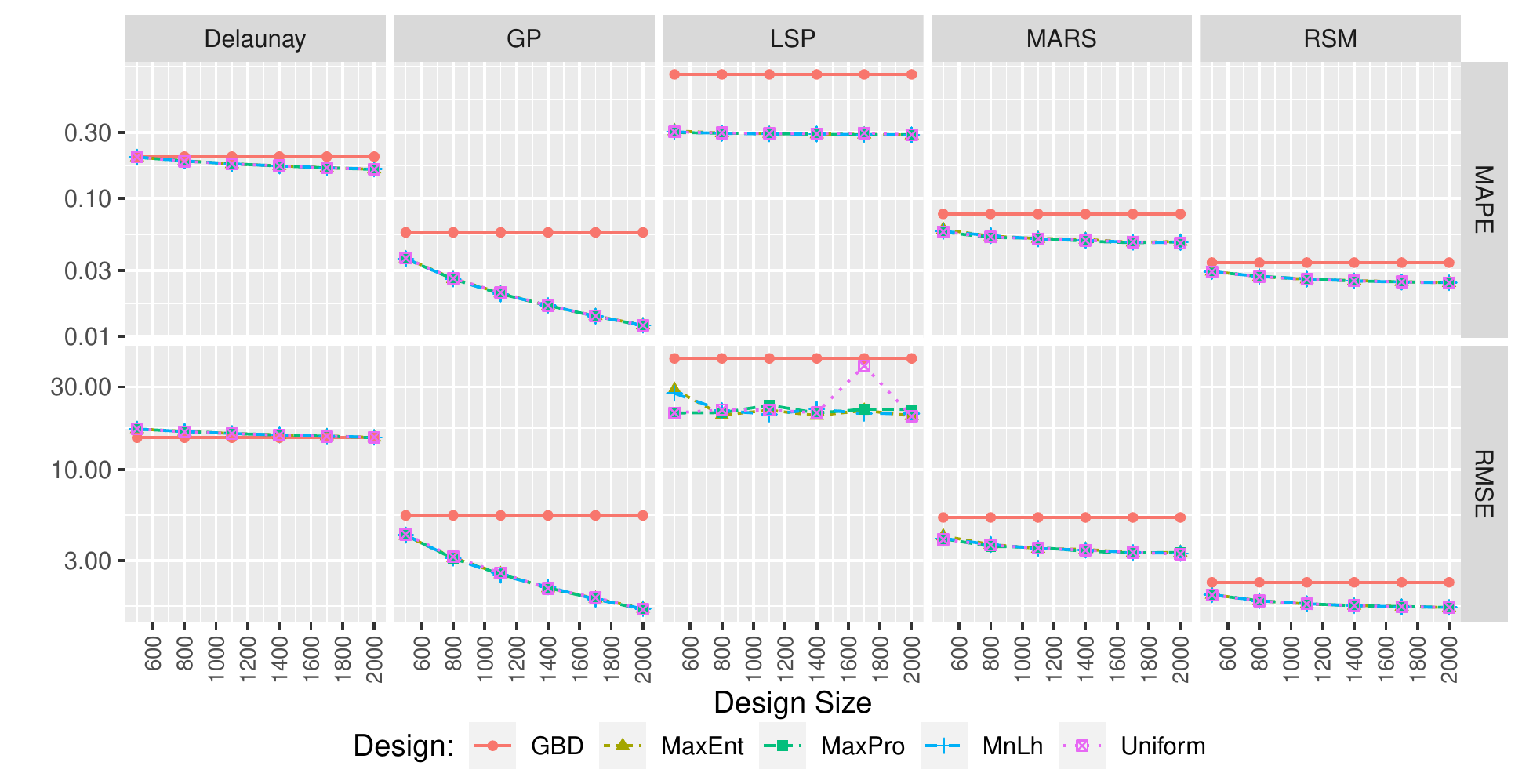}
\caption{Plot of RMSE and MAPE as functions of design size under the Borehole test function. Here, the GBD has a fixed size of $3^8 =6561$ due to the computational limit.}\label{fig:boresimulation}
\end{figure}

	For each test function, we plot the average RMSE and MAPE versus design size for each combination of approximation methods and design strategies. The results are shown in Figures~\ref{fig:colsimulation},~\ref{fig:friedsimulation}, and~\ref{fig:boresimulation}. The results show that the overall error decreases as the design size increases, but at different rates for different methods and problems. For the two four-dimensional test functions, we can see an interaction effect between the approximation method and design strategy. Under numerical approximation methods, GBD has a smaller prediction error as size increasing compared to the SFDs. However, under statistical models, the trend is the opposite. One possible explanation for this behavior is that GBDs have good geometric properties and both numerical approximation methods, Delaunay and linear Shepard, rely on geometric properties to make predictions. For the eight-dimensional Borehole function, the $3^8$ GBD is shown as a horizontal line in Figure~\ref{fig:boresimulation}. In general for the Borehole function, the GBD does not perform as well as the SFDs, despite the fact that each SFD has a smaller design size.

	\begin{table}
		\caption{Test set size $n_g$ and replication time $B$ in the synthetic data analyses.}
		\label{tbl:sim_pars_summary}
		\centering
		\begin{tabular}{c|rr|rr|rr}
			\hline\hline
			\multirow{2}{*}{Surrogate Model}&\multicolumn{2}{c|}{Friedman}&\multicolumn{2}{c|}{Colville}&\multicolumn{2}{c}{Borehole}\\
			\cline{2-7}
			& $n_g$ & $B$& $n_g$ & $B$& $n_g$ & $B$\\
			\hline
			RSM & 10000 & 30& 10000& 30 &5000&60\\
			Delaunay & 10000 & 30& 10000& 30 &5000&60\\
			LSP & 10000 & 30& 10000& 30 &5000&180\\
			MARS& 10000 & 30& 10000& 30 &5000&60\\	
			GP&10000 & 30& 10000& 30 &5000&60\\
			\hline\hline
		\end{tabular}
	\end{table}

\subsection{HPC Data Analysis}\label{sec:application}

In this section, we conduct a data analysis using the real data as described in Section \ref{sec:datacollection}.

\subsubsection{Model Fittings}\label{sec:realmodelfitting}

In HPC application, the data were collected using a full factorial design. Each of the four input factors has several unique levels and the throughput data were collected at each combination of levels. In total there are 2658 possible configurations. Since the real data itself is a GBD, we construct SFDs with sizes increased from 100 to 2700 with an increment of 200, and compare them to the real data. In this way, we investigate whether SFDs can achieve the same prediction accuracy as the GBD with a smaller design size.

A similar procedure as in Section~\ref{sec:simulation.procedure} is used. {Note that we did not consider the log transformation when building statistical models using the real data, although the response PVM is non-negative. Because the performance variability in our study is quite large, which is far away from 0. The responses are not highly skewed around zero. Thus, the non-negativity is of less concern. One possible side effect of the log transformation is that it can result in extremely large predictions after taking the anti-log to obtain the prediction in the original scale.}

There are a few modifications in the comparison procedure when analyzing the HPC data. Since the experiment region in the HPC performance variability problem is a discrete space, the SFD points are binned to the nearest feasible value after generation. Also because we do not know the true underlying surface in the real application as in the synthetic data analysis, we need to decide an underlying surface that can describe the real data well and also suitable for conducting comparisons. We choose to use a fitted model with methods in Section~\ref{sec:approximation.method} using the real data as the truth. In order to choose the most appropriate model, we first use a 10-fold cross validation (CV) to see the average CV prediction errors for different models. The result is summarized in Table~\ref{tbl:5modelscv}. Choosing the model that has the smallest CV prediction error is a natural idea. In this case, it is either Delaunay which has the smallest MAPE or GP with the smallest RMSE.
However, one property of Delaunay method is that the fitted model traverses every training data point. This means if we use the fitted Delaunay from the real data as the underlying truth, the response value we take from the fitted surface for the GBD, is exactly the real collected PVM. This benefits the GBD since the GBD has exactly the same data we used to build the true surface. When building approximation models to make predictions, the GBD has advantages over SFDs. Especially if we use the Delaunay approximation method, we will obtain 0 prediction error for the GBD. In this way, we can not make a fair comparison for the performance of the GBD and the SFD. If we consider the GP model, without the nugget effect, the GP model also goes through every training point, resulting in the same situation as using the Delaunay method. Even though in our analysis, we include the nugget effect so that the model does not go through every training point, and the GBD will not benefit as much as in the above situation. Using the fitted GP model is still not a fair comparison. That is because the variance-covariance matrix of the fitted GP model trained with the real data is featured by the evenly spaced design points. Potentially, using fitted GP model as the truth also benefits the GBD,  which has good geometric properties. Similarly, LSP model also has the same problem as the Delaunay.
Therefore, in order to have a fair comparison, we decided to use the MARS fitted model. It will not return exactly the same PVM for the GBD, also is not affected by the geometric property. Although its ability to describe the real data is not as good as the above three methods, we choose it as the truth in the real data analysis for a fair comparison.

\begin{table}
	\centering
	\caption{Average 10-folds CV error for the HPC data under different models.}
	\begin{tabular}{r|rrrrr}
		\hline\hline
		Model & RSM & MARS & Delaunay & LSP & GP \\
		\hline
		MAPE & 0.47 & 0.39 & 0.20 & 0.26 & 0.22 \\
		RMSE & 89476.99 & 80468.44 & 63394.17 & 76488.13 & 54854.81 \\
		\hline\hline
	\end{tabular}
	\label{tbl:5modelscv}
\end{table}

\begin{figure}
	
\centering
\includegraphics[width=1.0\textwidth]{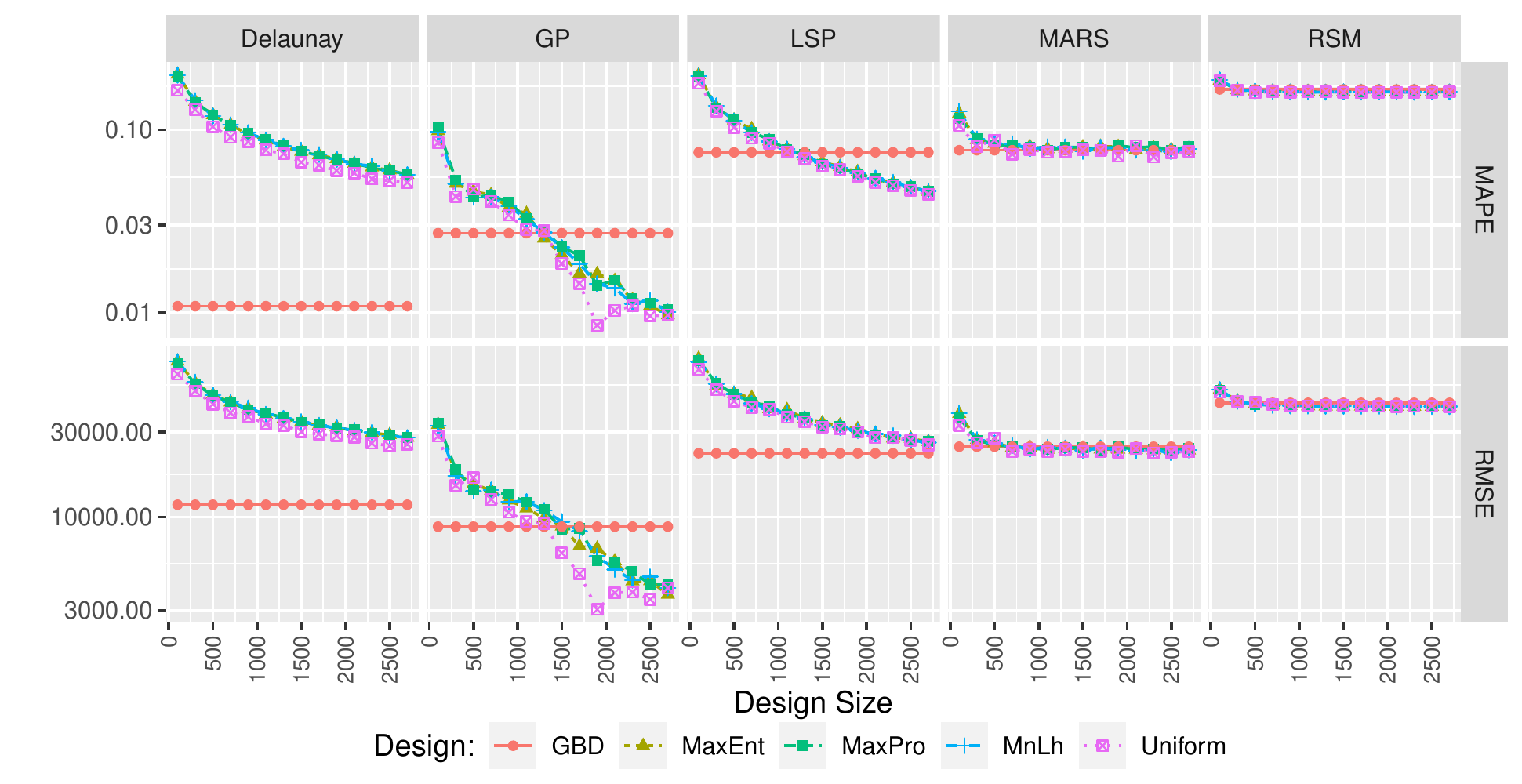}
\caption{Plot of RMSE and MAPE as functions of design size in the HPC application. The GBD is given by the locations where true data were collected. The true underlying surface is the one based on the MARS fit.}\label{fig:truemars_summary}
\end{figure}

\subsubsection{Comparisons}

Figure~\ref{fig:truemars_summary} describes the MAPE and RMSE of different methods and design combinations when using MARS fitted model as the underlying truth.  The error trends of all SFDs are similar to each other under every approximation method. GP with SFDs can achieve the smallest prediction error under both criteria. Like most cases in synthetic analysis, the GBD outperforms SFDs with Delaunay method. While  with GP, MARS and RSM models, SFDs can achieve the same prediction error as the GBD with smaller budget. For the LSP method, the SFDs outperform the GBD under MAPE criterion and are quite close with the GBD under the RMSE criterion.

One may notice that in Figures~\ref{fig:colsimulation} to~\ref{fig:truemars_summary}, the averaged prediction errors of SFDs are similar. One reason is that SFDs share similar properties, which are trying to fill out the whole design region to gain thorough information, so the average performance of various SFDs may perform similarly. Actually there do exist differences among different designs prediction ability if we look into each replication. However, these differences are hard to distinguish in the magnitude of those figures (i.e., if we do not zoom in).

\section{Conclusions and Recommendations}\label{sec:conclusion}
In this paper, comparisons are conducted for synthetic data analysis and HPC variability management application to explore the prediction accuracy of GBDs and SFDs under different approximation method. Overall the prediction error decreases as the design size increases. We find that no design outperforms all others uniformly. The GP with SFD, however, generates the best results under most scenarios. In previous work, the Delaunay method with GBD tends to have the smallest relative error~\shortcite{lux2018predictive}. In our analysis, with the maximum design budget, the best approximation method and design combination under each test function and real application is summarized in Table~\ref{tbl:summary_conclusion}. From this table we can see that GP outperforms other models under both error criteria. This is consistent with the CV prediction ability analysis with the real HPC data in Table~\ref{tbl:5modelscv}. {Besides, the GP model can quantify the prediction uncertainty, which allows a better understanding of the data and model.} Therefore, GP model is recommended when choosing the approximation method. If the data are collected in a grid-based manner, Delaunay method can also be considered. {However, as a numerical predictive model, the Delaunay method can not provide uncertainty quantifications at predictions.}

If we fix the approximation methods and look at the performance of designs, we notice that the GBD outperforms the SFD under the two numerical approximation methods (i.e., Delaunay and linear Shepard's method) for most cases. One possible reason is that GBDs have better geometric properties and these two methods depend on those properties. For the two statistical methods, MARS and RSM methods, SFDs have higher prediction accuracy compared to the GBD. One thing interesting with these two statistical methods is that, for SFDs, the increase of design size brings small improvement in prediction accuracy after the design size exceeds a certain value. That is, with MARS and RSM method, the prediction accuracy does benefit from the increase of design size in the early stage but after the design size achieves a relatively small budget, increasing design size can not guarantee a decrease in prediction error. While in contrast, under the GP surrogate, increasing the design size of SFDs always results in an obvious decrease in prediction errors.

With the best approximation method for each scenario, the design budget (i.e., the number of design points) of SFDs and GBDs to achieve the same or higher prediction accuracy are summarized in Table~\ref{tbl:summary_designsizereduction}.  We allow the design type that has weaker performance to take the maximum design budget in our analysis, and see how many runs the other type needs to have the same or higher precision. Then we can obtain Table~\ref{tbl:summary_designsizereduction} by comparing the error trends from Figures~\ref{fig:colsimulation} to~\ref{fig:truemars_summary}. When making comparisons, we only consider the design sizes that we computed the RMSE and MAPE in our analysis. For example, under the four-dimensional test functions, the design sizes of GBDs and SFDs take values from $54, 160, 375, 756, 1372$ (i.e., the sizes of GBDs with design budgets $3^4 = 81, 4^4 = 256, \ldots, 7^4 = 2401$ after adjusting for the constraint $x_3\geq x_4$). So when we fix the design type that has weaker performance at the maximum design budget $1372$, we only consider the prediction accuracy of the other type with size $54, 160, \ldots, 1372$. Similarly for Borehole function and the HPC application.
From Table~\ref{tbl:summary_designsizereduction} we can see that by choosing the right design type, we can save a large amount of time and cost. For the cases where SFDs are better, the number of design points needed for SFDs is about half of or less than that of the GBD to achieve the same prediction accuracy. Under the Friedman function, where the underlying surface is quite smooth, we observed that the GBD can also beat SFDs. However, GBD is not scalable to high-dimensional experiment regions. When the experiment region is of high dimension, the size of the GBD increases exponentially and building prediction models based on GBD will be time consuming even if we only consider a few unique points in each dimension. In practice, we can not guarantee that the underlying mechanism is a smooth function. So SFDs that are able to accommodate high-dimension and non-smooth functions are recommended when large numbers of input factors need to be considered in the model.

Despite the non-scalability of GBDs, they do have computational advantages when building models. Some matrix operations techniques can be used for GBDs because of their regular structure. To see the computational cost in our numerical studies, we report the running time for various designs to construct and build different models in the supplementary material. We notice that GBD almost costs no time to construct, while the construction time of SFDs increases with the size increases. Except for RSM, which requires very little computational resource, GBDs do perform with advantages compared to SFDs with the same design size in building predictive models across all other four models.

In the future, it will be interesting to investigate designs that can maximize the prediction accuracy based on a good choice of surrogates in the HPC setting. For example, G-optimal designs aim to minimize the maximum element on the diagonal of the hat matrix, which has the effect of minimizing the maximum variance among the predicted values. V-optimal designs minimize the average prediction variance among a set of points. G-/V-optimal designs could be considered because they minimize variance predictions. Another direction is that in our study, we used SFDs for continuous inputs and using a heuristic way to exclude points that do not satisfy the application constraint. A more refined approach that can propose discretized designs that maintain the space filling properties and also meet the constraint are desirable for solving real application.
One further step after obtaining desirable designs can be determining the system configuration that optimizes the HPC performance. For example, in \shortciteN{lixuoptimize} work, the optimal system configuration is determined as the configuration that can minimize the HPC variability while maintaining the HPC performance (i.e., the computing speed). This can provide insights in choosing system configurations in real HPC applications.
\begin{table}
\caption{Best design and approximation method combination with different test functions under the two criteria.}\label{tbl:summary_conclusion}
		\centering
		\begin{tabular}{c|c|c|c|c}
			\hline\hline
			\multirow{2}{*}{Test Function}&\multicolumn{2}{c|}{RMSE}&\multicolumn{2}{c}{MAPE}\\
			\cline{2-5}
			& Best Method & Best Design& Best Method & Best Design\\
			\hline
			Colville & GP & SFD&GP & SFD\\
			Friedman & GP& GBD&GP& GBD\\
			Borehole & GP & SFD&GP & SFD\\
			HPC application& GP & SFD&GP&SFD\\
			\hline\hline
		\end{tabular}
\end{table}

\begin{table}
	\caption{Design budget (i.e., the number of design points) of the GBD and SFDs to achieve a certain precision under both criteria.}\label{tbl:summary_designsizereduction}
	\centering
	\begin{tabular}{c|c|c|c|c}
		\hline\hline
		\multirow{2}{*}{Test Function and Method}&\multicolumn{2}{c|}{RMSE}&\multicolumn{2}{c}{MAPE}\\
		\cline{2-5}
		& SFD & GBD& SFD & GBD\\
		\hline
		Colville with GP& 54 & 1372& 375 & 1372\\
		Friedman with GP& 1372& 375&1372& 375\\
		Borehole with GP& 500 & 6561&500 & 6561\\
	HPC application with GP& 1500 & 2658&1300&2658\\
		\hline\hline
	\end{tabular}
\end{table}

\section*{Supplementary Materials}
The following supplementary materials are available online.
\begin{description}
\item[Additional Results:] The computing time of different designs and models (pdf file).
\item[Code and data:] Computing codes for the synthetic and real data analyses.  The HPC data are also included (zip file).
\end{description}

\section*{Acknowledgments}
The authors thank the editor, associate editor, and two referees, for their valuable comments that helped in improving the paper significantly.
The authors acknowledge Advanced Research Computing at Virginia Tech for providing computational resources and technical support that have contributed to the statistical results reported within this paper. The work is supported by funds from NSF under Grant CNS-1565314 and CNS-1838271 to Virginia Tech.

\section*{About the Authors}

Yueyao Wang is a Ph.D. Candidate in Department of Statistics at Virginia Tech. Her email address is yueyao94@vt.edu.

Li Xu is a Postdoctoral Researcher in Department of Epidemiology at Harvard University. His email address is li1992@vt.edu.

Yili Hong is Professor of Statistics at Virginia Tech. He is a member of ASQ. His email address is yilihong@vt.edu.

Rong Pan is Associate Professor, Program Chair of Industrial Engineering and Engineering Management at Arizona State University. He is a senior member of ASQ. His email address is Rong.Pan@asu.edu.

Tyler Chang is a Postdoctoral Appointee at Argonne National Laboratory. His email address is tchang@anl.gov.

Thomas Lux is a Senior Research Engineer at Amobee. His email address is tchlux@vt.edu.

Jon Bernard is a Ph.D. student in computer science at Virginia Tech. His email address is jobernar@vt.edu.

Layne Watson is Professor of Computer Science, Mathematics, and Aerospace and Ocean Engineering at Virginia Tech. His email is ltwatson@computer.org.

Kirk Cameron is Professor of Computer Science at Virginia Tech and Director of the Center for Computer Systems. His email is cameron@cs.vt.edu.


\end{document}